\documentclass[submission,copyright,creativecommons]{eptcs}
\providecommand{\event}{FVAV 2017} 
\usepackage{breakurl}             

\usepackage{amssymb}
\usepackage{amsthm}
\usepackage{amsmath}

\title{A Rational Agent Controlling an Autonomous Vehicle:
  \\ Implementation and Formal Verification}
\author{Lucas E. R. Fernandes \qquad Vinicius Custodio \qquad Gleifer V. Alves
\institute{Informatics Department,
UTFPR - Federal University of Technology - Paran\'a - Campus Ponta Grossa\\ Ponta Grossa, Brazil}
\email{\{lucfer, viniciuscustodio\}@alunos.utfpr.edu.br  \qquad gleifer@utfpr.edu.br}
\and Michael Fisher 
\institute{Department of Computer Science, University of Liverpool, Liverpool, United Kingdom}
\email{MFisher@liverpool.ac.uk}
}
\def\titlerunning{A Rational Agent Controlling an Autonomous Vehicle}
\def\authorrunning{L. Fernandes, V. Custodio, G. Alves \& M. Fisher}
\def\copyrightholders{L. Fernandes, V. Custodio, G. Alves \& M. Fisher}

\begin{document}
\maketitle

\begin{abstract}
The development and deployment of Autonomous Vehicles (\textsf{AVs})
on our roads is not only realistic in the near future but can also
bring significant benefits. In particular, it can potentially solve
several problems relating to vehicles and traffic, for instance: (i)
possible reduction of traffic congestion, with the consequence of
improved fuel economy and reduced driver inactivity; (ii) possible
reduction in the number of accidents, assuming that an \textsf{AV} can
minimise the human errors that often cause traffic accidents; and
(iii) increased ease of parking, especially when one considers the
potential for \emph{shared} AVs. In order to deploy an \textsf{AV}
there are significant steps that must be completed in terms of
hardware and software. As expected, software components play a key
role in the complex \textsf{AV} system and so, at least for safety, we
should assess the correctness of these components. 

In this paper, we are concerned with the high-level software
component(s) responsible for the decisions in an \textsf{AV}. We
intend to model an \textsf{AV} capable of navigation; obstacle
avoidance; obstacle selection (when a crash is unavoidable) and
vehicle recovery, etc, using a rational agent. To achieve this, we
have established the following stages. First, the agent plans and
actions have been \emph{implemented} within the \textsc{Gwendolen}
agent programming language. Second, we have \emph{built} a simulated
automotive environment in the \texttt{Java} language. Third, we have
formally \emph{specified} some of the required agent properties
through LTL formulae, which are then formally verified with the
\textsf{AJPF} verification tool. Finally, within the \textsf{MCAPL}
framework (which comprises all the tools used in previous stages) we
have obtained formal verification of our \textsf{AV} agent in terms of
its specific behaviours. For example, the agent plans responsible for
selecting an obstacle with low potential damage, instead of a higher
damage obstacle (when possible) can be formally verified within
\textsf{MCAPL}. We must emphasise that the major goal (of our present
approach) lies in the formal verification of agent plans, rather than
evaluating real-world applications. For this reason we utilised a
simple matrix representation concerning the environment used by our
agent.
\end{abstract}


\section{Introduction}

As society advances, new technologies emerge that can provide new
methods for helping people in their daily routine, as well as renewing
seemingly obsolete objects via technology allowing society to retain
benefits from the object while removing some if its problems. An
object that is going through a renovation is the road vehicle. Such
vehicles are very useful on a daily basis, allowing people to move
around quickly and cheaply. However, in recent years the process of
urbanisation has brought a huge amount of people into cities and,
consequently, has vastly increased the population of cars on our
roads~\cite{united_nations_2015}. This growth has resulted in
significant problems, such as traffic
congestion~\cite{fagnant_kockelman_2013}, increased occurrence of
accidents~\cite{peden_2004}, and significant additional pollution
emitted by such vehicles.

Autonomous vehicles represent a revolution for the automotive
industry~\cite{wallace_2012}, not only involving new types of vehicles
but also promising solutions to some of the above problems. Autonomous
Vehicles, i.e: vehicles with the ability to drive themselves, will
allow society to utilise the potential for road vehicles to their
maximum. Usually a person has their own car, meaning that it will
often only be used for a short period of time and will be left parked
for the rest of the time. Cliffe~\cite{cliffe:2016} asserts that
nowadays cars are used only 5\% of the time, but with the advent of
autonomous vehicles, this number is expected to increase to
75\%. According to~\cite{mit_media_lab}, each US citizen spends around
250 (typically, unproductive) hours inside a car and 40\% of the fuel
on urban journeys is used while searching for parking spaces.

\subsection{Autonomy and Intelligence $\longrightarrow$ Rational Agents}
In order to develop an effective and efficient \textsf{AV}, it is
necessary to endow the vehicle with some form `intelligence' that will
allow it not only to perceive the world but to react to it
appropriately. One way to provide this intelligence to a vehicle is
through the use of \emph{intelligent agents}. In the context of
information systems, agents are entities that have the capacity to
work independently and often, to achieve their own
goals~\cite{wooldridge_1995}. \emph{Reactive} agents are devised so as
to promptly respond to changes in their environment. These agents have
many similarities to adaptive systems, with environmental changes
being the main (sometimes only) trigger for change in
behaviour. \emph{Rational} agents are often more sophisticated
entities, capable of taking the initiative, planning and acting
autonomously, and doing so (rationally)in line with their own goals
(as well as environmental constraints). In practical autonomous
systems, a combination of such elements is often used. Reactive or
adaptive components deal with fast environmental interaction, while a
rational agent deals with the high-level
`intelligence'~\cite{LVDFL10:ALCOSP}. Such a combination has several
advantages, not only in terms of combining efficient interaction with
high-level reasoning (as above), but also providing a central
component (the rational agent) where the decision-making and
responsibility are encapsulated. This then presents the possibility of
deeper analysis of the agent's choices, for example through formal
verification~\cite{CACM13}.

\subsection{Reliability and Correctness $\longrightarrow$ Formal Verification}
A ratonal agent is a consistent way to endow a system with
intelligence. Nevertheless, a vehicle works in critical situations
that can put lives in risk. The National Highway Traffic Safety
Administration (NHTSA)~\cite{nhtsa_2012}, a U.S. agency responsible
for regulating autonomous vehicles, advocates that the level of
intelligence of the vehicles has to surpass human driving abilities
and so it is expected that the number of accidents caused by human
factors drops towards zero in the era of autonomous (also known as
\emph{self-driving}) cars.  As mentioned
in~\cite{mayer-schonberger_last_nodate}, the Waymo project (formerly
known as \textit{Google Self-Driving Car}) has decreased the number of
human interventions on its road tests --- these have fallen from 0.8
to 0.2 interventions per thousand miles between 2015 and 2016. As a
consequence, one may infer that reduced human intervention will
(likely) mean a safer \textsf{AV}. So, considering the deployment of
an \textsf{AV} with its actions and decisions controlled by hardware
and (particularly) software mechanisms, one may wonder: \emph{how can
  we guarantee that such a complex system will work reliably?}

An \textsf{AV} has a control architecture comprising both low-level
and high-level components. The low-level components include sensors
(and actuators) for object recognition, mapping, navigation, vehicle
direction, obstacle avoidance,
etc~\cite{DBLP:journals/corr/UlbrichRREBDNM17}, while the high-level
components are often responsible for the decisions concerning which
low-level components should be used and also when they must be
activated.  Assuring that a given (complex, computational) system (for
example, an \textsf{AV}) is able to always act safely and achieve its
goals thoroughly is a very hard problem. The construction of an
\textsf{AV} requires several different analysis mechanisms, not only
for hardware but also for software development, especially as software
components can range among different layers throughout the deployment
of an \textsf{AV}. One approach that seeks strong correctness of a
system is \emph{formal verification}. This involves providing a
mathematical (usually logical) proof that the system matches some
formal requirement. This is quite a strong commitment and, while
formal verification can potentially be applied to assess many
different layers within the architecture, we will primarily use it to
prove that the central rational agent makes the correct
decisions~\cite{CACM13}.

\subsection{Paper Overview}
We here intend to construct a rational agent in order to control the
high-level decisions within an \textsf{AV}. As we do not construct a
real \textsf{AV}, we will enbed a single rational agent within an
\textsf{AV} simulation environment. This set of decisions the agent
can take is represented as the possible plans for the agent. In this
study, the following plans have been defined: \emph{vehicle
  navigation}; \emph{obstacle avoidance}; \emph{obstacle selection}
(when a crash is unavoidable, the agent shall choose the obstacle with
the lowest damage); and \emph{vehicle recovery} (when a crash occurs).
    
We are concerned with verifying the choices taken by our rational agent. For instance, 
\textit{does our agent} (based on its sensor input) \textit{always
   select the expected action and avoid an obstacle} (when possible)?
%

This is the kind of question that could properly answered through the
use of formal verification on the rational agent and, indeed, we will
use a formal verification framework for agent programming languages
called \emph{Model Checking Agent Programming Languages}
(MCAPL)~\cite{Dennis:2012}. Within MCAPL, the \textsc{Gwendolen}
language~\cite{Dennis_2008} is used for programming the rational agent
and then the \emph{Agent Java Path Finder} (\textsf{AJPF})
model-checker is applied to these programs in order to formally verify
agent properties~\cite{bordini_2008}. The \textsf{AJPF} checker
carries out an exhaustive analysis to assess whether, in all possible
executions, the agent's behaviour matches it specified requirement. If
not, a counter-example is generated.

The properties to be assessed represent the required actions and
choices that the agent is supposed to take, depending on its
beliefs. These agent beliefs are, in turn, obtained from a simulated
automotive environment built in the \texttt{Java} language (since the
\textsf{MCAPL} framework works with such a language). The environment
comprises routes in which the agent can move towards its
destination. These routes are limited (within the boundaries of the
environment) and may also involve obstacles. There are three different
varieties of obstacle according to the level of possible damage: low,
moderate or high. With this, we can define simple formal requirements
to describe \textsf{AV} properties, such as the one mentioned
above. These formal properties are written as LTL (linear temporal
logic) formulae in order to be formally verified via \textsf{AJPF}
tool.
        
The reminder of the paper is organised as follows. Section 2 presents
definitions concerning rational agents and formal verification. Next,
in Section 3, we describe the modelling and implementation of our
agent. Section 4 shows the corresponding formal verification, while
Section 5 provides concluding remarks.

\section{Rational Agents and Formal Verification}

In this section, we briefly describe some aspects used in this work.
Here we use the notion of a \emph{rational agent} provided by Rao and
Wooldridge~\cite{wooldridge_foundations_1999}. The authors establish
that rational agents are software entities that perceive their
environment through sensors. Besides, the agents have a model of their
environment, and they can reason about it. Then, based on its own
mental state, a rational agent can take actions that may change the
environment~\cite{wooldridge_foundations_1999}.  A rational agent can
be implemented in a number of ways, but we choose to utlise the BDI
(\textit{Belief, Desire and Intention}) architecture. In our work we
have used a BDI programming language, named
\textsc{Gwendolen}~\cite{Dennis_2008}. This programming language can
be used with the \textsf{MCAPL} framework, which comprises
\textsc{Gwendolen} and the linked \textsf{AJPF}
tool~\cite{Dennis:2012}.

\textsf{AJPF} is an extension of the \emph{Java Path Finder} (JPF)
tool, which is used for formal verification of \texttt{Java}
programs~\cite{VisserHBPL03}. The \textsf{AJPF}
system~\cite{bordini_2008} has been conceived to work with BDI agent
programming languages, where the agent code can be formally verified
against given properties. These properties are written in a
\emph{Property Specification Language} (\textsf{PSL}) which is itself
based on LTL. The syntax of \textsf{PSL} is defined as follows. The
syntax for property formulae, $\phi$, is given bellow, where: $ag$
refers to a specific agent in the system; $f$ is a first-order atomic
formula; and $\lozenge$ ("eventually") and $\square$ ("always") are
standard linear temporal logic operators~\cite{Fisher:Book:2011}.

\begin{center}
$\phi $ $:: =$ $ \mid \mathsf{B}_{ag} \ f \mid  \mathsf{G}_{ag} \ f \mid \mathsf{A}_{ag} \ f \mid \mathsf{I}_{ag} \ f \mid \mathsf{ID}_{ag} \ f \mid \mathsf{P}(f) \mid 
\phi \lor \phi \mid \phi \land \phi \mid \neg \phi \mid \phi \ \mathsf{U} \ \phi 
\mid \phi \ \mathsf{R} \ \phi \mid \lozenge \phi \mid \square \phi$
\end{center}

\noindent Here,     
    \begin{itemize}
    \item $\mathsf{B}_{ag} \ f$ is true if $ag$ believes $f$ to be true, 
        
    \item $\mathsf{G}_{ag} \ f$ is true if $ag$ has a goal to make $f$ true, 
        
        \item \emph{and so on} with $\mathsf{A}$ representing actions,
          $\mathsf{I}$ representing intentions, $\mathsf{ID}$
          representing the intention to take an action, and
          $\mathsf{P}$ representing percepts, \textit{i.e.},
          properties that are true in the environment.
    \end{itemize}


\section{Agent: Modelling and Implementation}

Our work brings a new approach for modelling a rational agent in order
to behave as controller for an autonomous vehicle, where a simple
driving mechanism is considered. As mentioned in Section~1, we 
focus on the high-level components of an \textsf{AV} control
architecture. The agent acts as a taxi performing rides, picking up
and dropping off passengers in different positions in its
environment. In order to do so, the vehicle can move in four different
directions, being able to avoid static and known obstacles
ahead. Given a situation when a collision is unavoidable, the agent
should try to ensure the least amount of physical damage to the
vehicle, whenever this is possible.

Answering to passengers requests for rides and attempting to complete
them is the agent's main, and initial, desire.  Information about such
rides is provided by the environment, which creates (randomly) a list
$P$ of passengers, where $|P| \; \in \mathbb{N}$, and $p$ is a given
passenger ($p \in P$). Each $p$ is assigned a specific ride $r$, where
$R$ is a list of rides and $r \in R$. Each $r$ has two
distinct positions in the environment: \textbf{starting}
point and \textbf{destination} point. The positioning in
the environment is a square matrix of positions $A$ of order $n$; thus
$A_{n, n} \;|\; $\texttt{$n \in \mathbb{N}$}. Each element in the
matrix is a coordinate identified by $(X,Y) \mid (X,Y) \subset A$,
where $X$ and $Y$ are, respectively, the corresponding row and column
indexes of that element in $A_{n, n}$. Therefore, $X$, $Y$ $\in$
$\mathbb{N} \;| \; 0 \leq X, Y < n $.  Furthermore, a coordinate
($X,Y$) may have an obstacle $\theta$ contained in itself, hence
$\theta \subset $ ($X,Y$). Since our environment is static, this
statement will be true during all of the agent's activity. We
highlight the agent does not have \emph{complete} knowledge of
its environment. It will acquire further knowledge as it explores the
environment.

There are four sets of plans implemented with nine possible actions
over the environment responsible for determining the agent's
behaviour.

The first set of plans focusses on providing the agent abilities to
perform a sequence of rides. As its first priority, the agent gathers
its position in the environmental matrix by using the action
\texttt{localize} from the first set of plans, once there are no
unprocessed beliefs in the agent's belief set. Rides are requested
from the environment by the agent using an action \texttt{get\_ride},
which provides the coordinates of the starting and destination
points of any new ride or the information that there are no passengers left;
when the latter happens, the agent stops its
execution. 

For every new ride, the agent attempts to move the vehicle from its
current location to the ride's starting point. Similarly, after that the agent proceeds to attain the
destination point to drop off a passenger and conclude its ride. In
both cases it is said the agent does a route, and on
concluding this, it parks with an action \texttt{park} to allow the
passenger to get in and out of the vehicle. If, before starting a ride
or in any given moment within the ride, the agent believes that there
is an obstacle in either the ``pick up'' or ``drop off'' location,
that ride will be refused through an action \texttt{refuse\_ride}.

The agent asks the environment the location of the destination point 
of its current route is, in relation to the vehicle's current
coordinate. This is achieved by invoking an action \texttt{compass}.
In order to complete the routes, the agent is able to move in four
directions with an action \texttt{drive}, with these movements
relating to the way  the vehicle is maneuvering in the matrix. Therefore, a movement in a row, either increasing or
decreasing a coordinate's indices, is said to be moving, respectively
\textit{north} and \textit{south}. Meanwhile, movements in the
matrix's column that increase or decrease its coordinate's indexes are
denoted as, respectively, \textit{east} and \textit{west}.

When the agent moves to a new position in the environment, it receives
information about coordinates in its
surroundings, where such coordinates are strictly one movement away from the agent's current position. The
information received from the environment tells the agent about the
existence or not of any obstacle, as well as which direction from the
agent's current location will lead to the obstacle. As previously stated, 
the environment is static and thus the agent
will store beliefs related to obstacles during all its activity.

Each position through which the vehicle travels along a route stores
information that helps the agent determine if a particular direction
will be useful or not in reaching its destination. In this process,
the agent gathers belief referring which directions led to a
destination coordinate, as well as directions are taken to leave that
position. If, during  a route there is a direction
\textit{d1} the agent moved from a coordinate (X,Y) towards a
destination and an opposite direction \textit{d2} that the vehicle got
to the same location (X,Y), it is said that in order to reach the
current destination, the agent can \emph{not} move towards \textit{d1}
from (X,Y). This is necessary to avoid duplicate movements during the
same path. Thus, as it moves, the agent will generate a (partial)
model of the environment where it is located and will discern which
directions from its current position are invalid for completing a
route, determining the failed attempts to complete a route. The
situations detected are called 'known routes' and these routes should be
avoidable by the agent. 

During a route, if the agent finds an obstacle in the direction of its movement or known route, a set of obstacle
avoidance plans is used. This set gives priority to convenient route
adaptations taking the \textsf{AV} towards the destination of the
current route. If the agent finds an obstacle while moving through
rows of the location matrix and, in order to achieve its destination,
the vehicle will now need to move through many columns in the future,
the agent will adapt its route maneuvering towards the destination's
column until one move in a row can be undertaken. Similarly, if the
agent is moving between columns and it encounters an obstacle and it
believes that the current destination, in the past, required movements
through columns, then a maneuver in such a direction is executed until
the agent can directly move towards the destination. 

If no movement is possible because the destination is strictly in one
direction, the agent will instead opt to move towards an available
column (in case it is moving between rows) or row (in case it is
moving between columns).  Whenever a route adaptation fails, the agent
will turn the vehicle to the opposite direction. If the vehicle is
unable to adapt its route in three alternative directions, regardless
of whether any of these have been previously chosen as routes, the
agent will then move in that direction where movement is possible,
when said situation is interpreted as a 'dead end'. In the last two
situations, the agent will add beliefs through an action
\texttt{no\_further\_from} attaching the coordinate that proved
invalid to reach the destination, so the positions surrounding will
recognise it as part of a 'known route'.

As discussed earlier, there are scenarios where the collision of the
vehicle is unavoidable from its current position; such situations
have, as a precondition, the existence of three obstacles surrounding
the agent. Hence, if the vehicle moves in any of the three different
directions, a crash will occur. In the case of a direction not
containing an obstacle but the vehicle, the environment considers it
to be temporarily obstructed and therefore, the agent has no maneuver
options in which there will be no collision.  Once the formal
precondition can be verified as true, the environment generates a
probability $T$ that, in its next move, the vehicle will collide with
an obstacle; likewise, there is a probability $F$ that the agent can
maneuver and avoid the obstacle, where $T, F \in \mathbb {R} \;|\; 0
\leqslant T, F \leqslant 1 $ and $ T + F = 1$. This is associated with
a damage level for each obstacle. The discrete possible
classifications to be assigned to an obstacle are \textit{low},
\textit{moderate} and \textit{high}, where for each of them contains a
probability $L$, $M$ and $H$ of collision occurring, where $L, M, H
\in \mathbb{R} \;|\; 0 \leqslant L, M, H \leqslant 1 $ and $L + M + H
= 1$.  In such situations, the agent will seek to choose to move
towards an obstacle whose collision will cause the least
amount of physical damage to the vehicle. 



\section{Formal Verification of Agent Plans}
In accordance with the previous actions and plans defined for the
agent, we have carried out formal verification by using the
\textsf{AJPF} tool. Here we will only present the verification that
represents the scenario responsible for obstacle selection, but all
agent's plans have been formally verified\footnote{Plans, models, and
  verifications are available at \url{https://github.com/AVIA-UTFPR/SAE}}.  We have chosen to
describe verificaiton of the obstacle selection plan since it
potentially has the most negative effect on the agent's desired
behaviour.  For instance, an undesired behaviour (when an obstacle
selection takes place) may lead to severe car damage.  This property
ensures that the agent will choose to move to some coordinate where
the damage is minimal in case of an unavoidable collision.

The \textbf{property} to be verified is: 
''During the entire agent's activity, in the occurrence of an
    unavoidable collision where there is at least one direction that
    the agent moves towards that will give a \textit{low}
    level of damage to the vehicle, then that will be the direction
    chosen by the agent, \textbf{and} in the case when there is
    \textbf{not} a direction where the level of damage is \textit{low}
    but there is one where a collision will be \textit{moderate}, the
    agent will move towards that direction''. 

The verification of the such property is presented in
Table~\ref{quadro:coc}, where it is formally specified by a LTL
formula\footnote{The specification of the machine used to perform this
  verification is as it follows: MacOS Sierra 10.12.5 on a Macbook Pro
  (Early 2015) with Intel Core i5 2.7 Ghz processor and 8 GB of memory
  RAM.}. Note that, by using properties of \textsc{Gwendolen} and \textsf{AJPF} tool, it is possible to verify every
single possible outcome of obstacle classification.

\begin{table}[ht]
	\centering
	\caption{Unavoidable collisions and the choice for the least amount of damage}
	\begin{tabular}{|p{7.5cm}|p{7.5cm}|}
 		\hline \hline
 			\textbf{Scenario settings} & \textbf{Value} \\
 		\hline
			Agent's initial Position &  \texttt{(2,1)} \\
 		\hline
			Total of rides & 1 \\
 		\hline
			 Rides' Coordinates & \{ (\texttt{(4,1)}-\texttt{(4,0)}) \} \\
 		\hline
			  Total of  obstacles & 3 \\
 		\hline
			 Obstacles' Coordinates & \{\texttt{(1,1)}, \texttt{(2,2)}, \texttt{(2,0)}\} \\
 		\hline\hline
 			\multicolumn{2}{|c|}{ \textbf{Formal Verification by AJPF} } \\
 		\hline\hline
 			\multicolumn{2}{|p{15.6cm}|}{\begin{flushleft}$ \Box ((\lozenge (B_{vehicle}unavoidable\_collision(\_,\_) \; \wedge \;  B_{vehicle}obstacle\_damage(\_,\_,\_,low) ) \implies \lozenge G_{vehicle}colide\_obstacle(\_, low) ) \wedge (\lozenge (B_{vehicle}unavoidable\_collision(\_,\_) \; \wedge \;  \sim B_{vehicle}obstacle\_damage(\_,\_,\_,low) \; \wedge \;  B_{vehicle}obstacle\_damage(\_,\_,\_,moderate) ) \implies \lozenge G_{vehicle}colide\_obstacle(\_, moderate) ) )$ 
 			\end{flushleft}} \\
	 	\hline
 			\textbf{Property} & \textbf{Feedback} \\
 		\hline
			 Total of environment states & 25465 \\
 		\hline
			 Maximum search depth & 444 \\
 		\hline
			 Total of states B\"uchi automata created by AJPF  & 12 \\
 		\hline
			 Total of run-time & 01:09:58\\
 		\hline
			 Total of RAM memory used & 3866MB\\
		\hline
	\end{tabular}
	\label{quadro:coc}
\end{table}

\section{Conclusion}

In this paper, we have presented a \textsc{Gwendolen} agent with plans
and actions devised to model simple, basic high-level components of an
Autonomous Vehicle. One of these components involves the handling of
static obstacles, which works in a twofold manner: (i) obstacle
avoidance; and (ii) obstacle selection (\textit{when a crash is
  unavoidable}). After that, we have deployed a simple simulated
environment in  \texttt{Java} language with the corresponding
scenario in order to test the plans and actions designed for our
agent. This environment represents obstacles as well as the routes
traced by the agent. Next, we have chosen formal properties concerning
the desired behaviour of our agent, specifically when taking into
account static obstacles. These properties have been
formally specified as LTL formulae and verified with the \textsf{AJPF}
tool. As a result, we have successfully applied the \textsf{MCAPL}
framework in order to obtain the modelling, implementation and formal
verification of a simple rational agent within an \textsf{AV}.

This work is part of  a project named \textit{Autonomous
  Vehicles with Intelligent Agents} (AVIA). And here we present 
  the very first results of AVIA, where we are
not (yet) concerned with real world interactions. That is the reason
why we have a minimal set of options and have used a (simple) matrix
representation in the environment used by the rational agent.
In future work, we shall extend the representation for handling
multiple traffic lanes and dynamic obstacles in order to fit in more
real world challenges and applications. Within this, we will define
the implementation of plans and actions to avoid the dynamic
obstacles. Furthermore, we aim to improve our simulated environment
with the representation of lanes and roads so that we can define plans
for lane crossing, lane departure and lane keeping.


\textbf{Acknowledgments.} The work of the last author was 
supported by EPSRC through its \emph{Verifiable Autonomy} research
project, EP/L024845.


\nocite{*} 
\bibliographystyle{eptcs} 
\bibliography{refers-fvav-paper}

\begin{thebibliography}{10}
\providecommand{\bibitemdeclare}[2]{}
\providecommand{\surnamestart}{}
\providecommand{\surnameend}{}
\providecommand{\urlprefix}{Available at }
\providecommand{\url}[1]{\texttt{#1}}
\providecommand{\href}[2]{\texttt{#2}}
\providecommand{\urlalt}[2]{\href{#1}{#2}}
\providecommand{\doi}[1]{doi:\urlalt{http://dx.doi.org/#1}{#1}}
\providecommand{\bibinfo}[2]{#2}

\bibitemdeclare{inproceedings}{bordini_2008}
\bibitem{bordini_2008}
\bibinfo{author}{Rafael~H. \surnamestart Bordini\surnameend},
  \bibinfo{author}{Louise~A. \surnamestart Dennis\surnameend},
  \bibinfo{author}{Berndt. \surnamestart Farwer\surnameend} \&
  \bibinfo{author}{Michael \surnamestart Fisher\surnameend}
  (\bibinfo{year}{2008}): \emph{\bibinfo{title}{Automated Verification of
  Multi-Agent Programs}}.
\newblock In: {\sl \bibinfo{booktitle}{Proceedings of the 2008 23rd IEEE/ACM
  International Conference on Automated Software Engineering}},
  \bibinfo{series}{ASE '08}, \bibinfo{publisher}{IEEE Computer Society},
  \bibinfo{address}{Washington, DC, USA}, pp. \bibinfo{pages}{69--78}.
\newblock \urlprefix\url{http://dx.doi.org/10.1109/ASE.2008.17}.

\bibitemdeclare{misc}{cliffe:2016}
\bibitem{cliffe:2016}
\bibinfo{author}{Mark \surnamestart Cliff\surnameend} (\bibinfo{year}{2016}):
  \emph{\bibinfo{title}{Driverless cars --- the route to more than smart
  cities}}.
\newblock
  \urlprefix\url{https://ingworld.ing.com/en/2016-1Q/12-column-m-cliffe}.
\newblock \bibinfo{note}{Accessed, 22 May 2017}.

\bibitemdeclare{inproceedings}{Dennis_2008}
\bibitem{Dennis_2008}
\bibinfo{author}{Louise~A. \surnamestart Dennis\surnameend} \&
  \bibinfo{author}{Berndt \surnamestart Farwer\surnameend}
  (\bibinfo{year}{2008}): \emph{\bibinfo{title}{Gwendolen: A BDI Language for
  Verifiable Agents}}.
\newblock In: {\sl \bibinfo{booktitle}{AISB Convention, University of
  Aberdeen}}, \doi{10.1.1.141.1549}.

\bibitemdeclare{article}{Dennis:2012}
\bibitem{Dennis:2012}
\bibinfo{author}{Louise~A. \surnamestart Dennis\surnameend},
  \bibinfo{author}{Michael \surnamestart Fisher\surnameend},
  \bibinfo{author}{Matthew~P. \surnamestart Webster\surnameend} \&
  \bibinfo{author}{Rafael~H. \surnamestart Bordini\surnameend}
  (\bibinfo{year}{2012}): \emph{\bibinfo{title}{{Model Checking Agent
  Programming Languages}}}.
\newblock {\sl \bibinfo{journal}{Automated Software Engineering}}
  \bibinfo{volume}{19}(\bibinfo{number}{1}), pp. \bibinfo{pages}{5--63},
  \doi{10.1007/s10515-011-0088-x}.
\newblock \urlprefix\url{https://doi.org/10.1007/s10515-011-0088-x}.

\bibitemdeclare{article}{fagnant_kockelman_2013}
\bibitem{fagnant_kockelman_2013}
\bibinfo{author}{Daniel~J. \surnamestart Fagnant\surnameend} \&
  \bibinfo{author}{Kara~M. \surnamestart Kockelman\surnameend}
  (\bibinfo{year}{2013}): \emph{\bibinfo{title}{Preparing a Nation for
  Autonomous Vehicles: Opportunities, Barriers and Policy Recommendations}}.
\newblock {\sl \bibinfo{journal}{Eno Center for Transportation}}
  \bibinfo{volume}{2}.

\bibitemdeclare{book}{Fisher:Book:2011}
\bibitem{Fisher:Book:2011}
\bibinfo{author}{Michael \surnamestart Fisher\surnameend}
  (\bibinfo{year}{2011}): \emph{\bibinfo{title}{An Introduction to Practical
  Formal Methods Using Temporal Logic}}.
\newblock \bibinfo{publisher}{Wiley}, \doi{10.1002/9781119991472}.
\newblock
  \urlprefix\url{http://eu.wiley.com/WileyCDA/WileyTitle/productCd-0470027886.html}.

\bibitemdeclare{article}{CACM13}
\bibitem{CACM13}
\bibinfo{author}{Michael \surnamestart Fisher\surnameend},
  \bibinfo{author}{Louise~A. \surnamestart Dennis\surnameend} \&
  \bibinfo{author}{Matthew \surnamestart Webster\surnameend}
  (\bibinfo{year}{2013}): \emph{\bibinfo{title}{{Verifying Autonomous
  Systems}}}.
\newblock {\sl \bibinfo{journal}{ACM Communications}}
  \bibinfo{volume}{56}(\bibinfo{number}{9}), pp. \bibinfo{pages}{84--93}.
\newblock \urlprefix\url{http://doi.acm.org/10.1145/2494558}.

\bibitemdeclare{inproceedings}{LVDFL10:ALCOSP}
\bibitem{LVDFL10:ALCOSP}
\bibinfo{author}{Nicholas \surnamestart Lincoln\surnameend},
  \bibinfo{author}{Sandor~M. \surnamestart Veres\surnameend},
  \bibinfo{author}{Louise~A. \surnamestart Dennis\surnameend},
  \bibinfo{author}{Michael \surnamestart Fisher\surnameend} \&
  \bibinfo{author}{Alexei \surnamestart Lisitsa\surnameend}
  (\bibinfo{year}{2010}): \emph{\bibinfo{title}{{An Agent Based Framework for
  Adaptive Control and Decision Making of Autonomous Vehicles}}}.
\newblock In \bibinfo{editor}{Erdal \surnamestart Kayacan\surnameend}, editor:
  {\sl \bibinfo{booktitle}{Proc. 10th {IFAC} International Workshop on the
  Adaptation and Learning in Control and Signal Processing, {ALCOSP} 2010}},
  \bibinfo{publisher}{International Federation of Automatic Control},
  \bibinfo{address}{Istanbul, Turkey}, pp. \bibinfo{pages}{310--317},
  \doi{10.3182/20100826-3-TR-4015.00058}.
\newblock \urlprefix\url{https://doi.org/10.3182/20100826-3-TR-4015.00058}.

\bibitemdeclare{misc}{mayer-schonberger_last_nodate}
\bibitem{mayer-schonberger_last_nodate}
\bibinfo{author}{Viktor \surnamestart Mayer-Schonberger\surnameend}
  (\bibinfo{year}{2017}): \emph{\bibinfo{title}{The last things that will make
  us uniquely human}}.
\newblock
  \urlprefix\url{http://www.bbc.com/future/story/20170309-the-last-things-that-will-make-us-uniquely-human}.
\newblock \bibinfo{note}{Accessed, 06 June 2017}.

\bibitemdeclare{misc}{mit_media_lab}
\bibitem{mit_media_lab}
\bibinfo{author}{\surnamestart {MIT Media Lab}\surnameend}
  (\bibinfo{year}{2016}): \urlprefix\url{https://www.media.mit.edu/}.
\newblock \bibinfo{note}{Accessed, 22 May 2017}.

\bibitemdeclare{book}{united_nations_2015}
\bibitem{united_nations_2015}
\bibinfo{author}{United \surnamestart Nations\surnameend}
  (\bibinfo{year}{2015}): \emph{\bibinfo{title}{Department of Economic and
  Social Affairs: Population Division. World Urbanization Prospects: The 2014
  Revision}}.
\newblock \bibinfo{publisher}{United Nations}.

\bibitemdeclare{misc}{nhtsa_2012}
\bibitem{nhtsa_2012}
\bibinfo{author}{\surnamestart NHTSA\surnameend} (\bibinfo{year}{2012}):
  \emph{\bibinfo{title}{Preliminary Statement of Policy Concerning Automated
  Vehicles}}.
\newblock
  \urlprefix\url{http://www.autoalliance.org/index.cfm?objectid=CC9678B0-A415-11E5-997E000C296BA163}.
\newblock \bibinfo{note}{Accessed, 22 May 2017}.

\bibitemdeclare{book}{peden_2004}
\bibitem{peden_2004}
\bibinfo{author}{Margie \surnamestart Peden\surnameend},
  \bibinfo{author}{Richard \surnamestart Scurfield\surnameend},
  \bibinfo{author}{David \surnamestart Sleet\surnameend},
  \bibinfo{author}{Dinesh \surnamestart Mohan\surnameend},
  \bibinfo{author}{Adnan~A. \surnamestart Hyder\surnameend},
  \bibinfo{author}{Eva \surnamestart Jarawan\surnameend} \&
  \bibinfo{author}{Colin \surnamestart Mathers\surnameend}
  (\bibinfo{year}{2004}): \emph{\bibinfo{title}{World report on road traffic
  injury prevention}}.
\newblock \bibinfo{publisher}{World Health Organization}.

\bibitemdeclare{article}{DBLP:journals/corr/UlbrichRREBDNM17}
\bibitem{DBLP:journals/corr/UlbrichRREBDNM17}
\bibinfo{author}{Simon \surnamestart Ulbrich\surnameend},
  \bibinfo{author}{Andreas \surnamestart Reschka\surnameend},
  \bibinfo{author}{Jens \surnamestart Rieken\surnameend},
  \bibinfo{author}{Susanne \surnamestart Ernst\surnameend},
  \bibinfo{author}{Gerrit \surnamestart Bagschik\surnameend},
  \bibinfo{author}{Frank \surnamestart Dierkes\surnameend},
  \bibinfo{author}{Marcus \surnamestart Nolte\surnameend} \&
  \bibinfo{author}{Markus \surnamestart Maurer\surnameend}
  (\bibinfo{year}{2017}): \emph{\bibinfo{title}{Towards a Functional System
  Architecture for Automated Vehicles}}.
\newblock {\sl \bibinfo{journal}{CoRR}} \bibinfo{volume}{abs/1703.08557}.
\newblock \urlprefix\url{http://arxiv.org/abs/1703.08557}.

\bibitemdeclare{article}{VisserHBPL03}
\bibitem{VisserHBPL03}
\bibinfo{author}{Willem \surnamestart Visser\surnameend},
  \bibinfo{author}{Klaus \surnamestart Havelund\surnameend},
  \bibinfo{author}{Guillaume~P. \surnamestart Brat\surnameend},
  \bibinfo{author}{Seungjoon \surnamestart Park\surnameend} \&
  \bibinfo{author}{Flavio \surnamestart Lerda\surnameend}
  (\bibinfo{year}{2003}): \emph{\bibinfo{title}{{Model Checking Programs}}}.
\newblock {\sl \bibinfo{journal}{Automated Software Engineering}}
  \bibinfo{volume}{10}(\bibinfo{number}{2}), pp. \bibinfo{pages}{203--232},
  \doi{10.1023/A:1022920129859}.

\bibitemdeclare{misc}{wallace_2012}
\bibitem{wallace_2012}
\bibinfo{author}{Richard \surnamestart Wallace\surnameend} \&
  \bibinfo{author}{Gary \surnamestart Silber\surnameend}
  (\bibinfo{year}{2012}): \emph{\bibinfo{title}{Self-Driving cars: The next
  revolution}}.
\newblock
  \urlprefix\url{https://assets.kpmg.com/content/dam/kpmg/pdf/2015/07/self-driving-cars-talkbook.pdf}.
\newblock \bibinfo{note}{Accessed, 23 May 2017}.

\bibitemdeclare{article}{wooldridge_1995}
\bibitem{wooldridge_1995}
\bibinfo{author}{Michael \surnamestart Wooldridge\surnameend} \&
  \bibinfo{author}{Nicholas~R. \surnamestart Jennings\surnameend}
  (\bibinfo{year}{1995}): \emph{\bibinfo{title}{{Intelligent Agents: Theory and
  Practice}}}.
\newblock {\sl \bibinfo{journal}{Knowledge Engineering Review}}
  \bibinfo{volume}{10}(\bibinfo{number}{2}), pp. \bibinfo{pages}{115--152},
  \doi{10.1017/S0269888900008122}.
\newblock \urlprefix\url{https://doi.org/10.1017/S0269888900008122}.

\bibitemdeclare{book}{wooldridge_foundations_1999}
\bibitem{wooldridge_foundations_1999}
\bibinfo{author}{Michael \surnamestart Wooldridge\surnameend} \&
  \bibinfo{author}{Anand \surnamestart Rao\surnameend} (\bibinfo{year}{1999}):
  \emph{\bibinfo{title}{Foundations of {Rational} {Agency}}}.
\newblock {\sl \bibinfo{series}{Applied {Logic}
  {Series}}}~\bibinfo{volume}{14}, \bibinfo{publisher}{Springer Netherlands},
  \doi{10.1007/978-94-015-9204-8}.
\newblock \urlprefix\url{http://www.springer.com/gp/book/9780792356011}.

\end{thebibliography}

\end{document}